\documentclass[conference]{IEEEtran}
\usepackage{amssymb,amsmath,amsthm}
\usepackage[lined,boxed,commentsnumbered,ruled]{algorithm2e}
\usepackage{algorithmic} 
\usepackage{graphicx}
\usepackage{stmaryrd}
\SetSymbolFont{stmry}{bold}{U}{stmry}{m}{n}
\usepackage{multirow,url}
\usepackage{color,cite}
\usepackage{cite}
\usepackage{tikz,pgf}
\usetikzlibrary{fadings,shapes,arrows,shadows}
\usepackage{makecell}
\usepackage{array}
\usepackage{overpic} 
\usepackage{subfigure}
\usepackage{url}

\usepackage{booktabs} 

\newtheorem{lemma}{Lemma}

\newtheorem{definition}{Definition}

\newtheorem{observation}{Observation}

\theoremstyle{plain}

\newcommand{\com}[1]{} 

\def\eq{\triangleq}

\def\G{\mathcal{G}}
\def\N{\mathcal{N}}     
\def\S{\mathcal{S}}     
\def\W{W}               
\def\u{u}               
\def\s{s}               
\def\bs{\boldsymbol{\s}} 

\def\T{\mathcal{T}}

\IEEEoverridecommandlockouts

\begin{document}

\title{Technical Report for ``User-Centric Participatory Sensing: A Game Theoretic Analysis'' \vspace{-3mm}}

\author{Xiaoyan Mo, Zhang Li, Lin Gao, Bin Cao, Tingting Zhang, and Tong Wang
\thanks{Authors are with the School of Electronic and Information Engineering, Harbin Institute of Technology, Shenzhen, China.
The first two authors  Xiaoyan Mo and Zhang Li  \textbf{contribute equally} to this work.
Lin Gao is the Corresponding Author.
Email: gaol@hit.edu.cn.}
\thanks{This work is supported by the National Natural Science Foundation of China  (Grant No. 61771162 and 61501211) and the Basic Research Project of Shenzhen (Grant No. JCYJ20160531192013063 and JCYJ20170307151148585).}
}
\maketitle

\addtolength{\abovedisplayskip}{-1mm}
\addtolength{\belowdisplayskip}{-1mm}

\begin{abstract}
Participatory sensing (PS) is a novel and promising
sensing network paradigm for achieving a flexible and scalable
sensing coverage with a low deploying cost, by encouraging
mobile users to participate and contribute their smartphones as
sensors. In this work, we consider a general PS system model with
\emph{location-dependent} and \emph{time-sensitive} tasks, which generalizes the
existing models in the literature. We focus on the task scheduling
in the \emph{user-centric} PS system, where each participating user will
make his individual task scheduling decision (including both the
task selection and the task execution order) distributively. Specifically,
we formulate the interaction of users as a strategic game
called \emph{Task Scheduling Game (TSG)} and perform a comprehensive
game-theoretic analysis. First, we prove that the proposed TSG
game is a potential game, which guarantees the existence of Nash
equilibrium (NE). Then, we analyze the efficiency loss and the
fairness index at the NE. Our analysis shows the efficiency at NE
may increase or decrease with the number of users, depending
on the level of competition. This implies that it is \emph{not} always
better to employ more users in the user-centric PS system, which
is important for the system designer to determine the optimal
number of users to be employed in a practical system.
\end{abstract}

\IEEEpeerreviewmaketitle

\section{Introduction}

\subsection{Background and Motivations}


With the development and proliferation of smartphones with rich build-in sensors and advanced computational capabilities,
we are witnessing a new sensing network paradigm known as \emph{participatory sensing (PS)} or \emph{mobile crowd sensing (MCS)}  \cite{bg1,bg2,bg3}, which relies  on the active participation and contribution of smartphone users (to contribute their smartphones as sensors).
Comparing with the traditional approach of deploying sensor nodes and sensor networks, this new sensing scheme can achieve a higher sensing coverage with a lower deployment  cost, and hence it better adapts to the  changing requirement of tasks and the varying environment.
Therefore, it has found a wide range of applications in environment, infrastructure, and community monitoring \cite{App-WeatherLah,App-OpenSignal, App-Atmos1,App-Waze}.

A typical PS framework often consists of (i) a service platform (\emph{server}) residing in the cloud and (ii)
a set of participating smartphone \emph{users} distributed and travelling on the ground \cite{bg1,bg2,bg3}.
The service platform launches many sensing tasks, possibly initiated by different requesters with different data requirements for different purposes;
and users subscribe to one or multiple task(s) and contribute their sensing data.
Due to the location-awareness and time-sensitivity of   tasks
and the geographical distribution of   users,
a proper \emph{scheduling} of tasks among users is critical for a PS system.
For example, if a task is scheduled to a user far away from its target location, the user may not be able to travel to the target location in time so as to complete the task successfully.

Depending on who (i.e., the server or each user) will make the task scheduling decision, there are two types of different PS models: \emph{Server-centric Participatory Sensing (SPS)} \cite{Luo-infocom2014, yang-mobicom12, Duan-infocom2012,Gao-2015,add1,add2}  and \emph{User-centric Participatory Sensing (UPS)} \cite{ups-0,ups-1,ups-2,ups-3,add3}. In the SPS model, the server will make the task scheduling decision and determine the joint scheduling of all tasks among all users, often in a centralized manner    with complete information (as in \cite{Luo-infocom2014, yang-mobicom12, Duan-infocom2012,Gao-2015,add1,add2}).
In the UPS model, each participating user will make his individual task scheduling decision and determine the tasks he is going to execute, often in a distributed manner with local information (as in\cite{ups-0,ups-1,ups-2,ups-3,add3}).
Clearly, the SPS model assigns more control to the server to make the (centralized) joint scheduling decision, hence can better satisfy the requirements of  various tasks.
The UPS model, however, distributes the control among the participating users and enables each user to make the (distributed) individual scheduling decision.
Hence, it can faster adapt to the varying environment and the changing requirement of individual users.~~~~~~~~~~~~~~~~~~~~~~~~~~~~~~~~~~~~~~~~~~~~~~~~~~

 {In this work, we focus on  \emph{the task scheduling in the UPS model}},
where the task scheduling decision is made by each user distributively.
Comparing with SPS, the UPS model has the appealing features of (i) low communication overhead and (ii) low computational complexity,
by distributing the complicated central control (and computation) among numerous participating users, hence it is more scalable.
Therefore, UPS is particularly suitable for a  fast changing environment (where the information exchange in SPS may become a heavy burden) and a large-scale system (where the centralized task scheduling in SPS may be too complicated to compute in real-time), and hence it has been adopted in
some commercial PS systems, such as Field Agent  \cite{example-1}
and Gigwalk \cite{example-2}.

\subsection{Related Work}

Many existing works have studied the task scheduling problem in different UPS models, aiming at either minimizing the energy consumption (e.g., \cite{add3, ups-1,ups-2}) or maximizing the social surplus (e.g., \cite{ups-0,ups-3}).
Specifically,
in \cite{add3}, Jiang \emph{et al.} studied the peer-to-peer based data sharing among users in mobile crowdsensing, but they considered neither the location-dependence nor the time-sensitivity  of tasks.
In \cite{ups-1}, Sheng \emph{et al.} studied the opportunistic energy-efficient collaborative sensing for location-dependent road information.
In \cite{ups-2}, Zhao \emph{et al.} studied the fair and energy-efficient
task scheduling in mobile crowdsensing with location-dependent tasks.
In \cite{ups-0}, He \emph{et al.} studied the social surplus maximization for location-dependent task scheduling in mobile crowdsensing.
However, the above works did not consider the time-sensitivity of tasks, where each task can be executed at any time.
In this work, we will consider both the location-dependence and the time-sensitivity of tasks.

Cheung \emph{et al.} in \cite{ups-3} studied the social surplus maximization scheduling for both location-dependent and {time-sensitive} tasks, where each task must be executed at a particular time.
Inspired by \cite{ups-3}, in this work, we will consider a more general task model, where each task can be executed within a \emph{valid time period} (instead of the particular time in \cite{ups-3}).
Clearly, \emph{our task model generalizes the existing models in  \cite{ups-0,ups-1,ups-2,ups-3,add3}}, as it will degenerate to the   models in \cite{ups-0,ups-1,ups-2,add3} by simply choosing an infinitely large valid time period for each task, and degenerate to the   model  in \cite{ups-3} by simply shrinking the valid time period of each task to a single point.~~~~~~~~~~~~~~~~~~~~~~~~~~~~~

\subsection{Solution and Contributions}

\begin{figure}
\vspace{-2mm}
	\centering
	\caption{An UPS Model with Location-Dependent Time-Sensitive Tasks. Each route denotes the task selection and execution order of each user.}
\label{fig:mcs-model}
\vspace{-3mm}
\end{figure}

In this work, we consider a general UPS model consisting of multiple tasks and multiple smartphone users, where each user will make his individual task scheduling decision distributively (e.g., deciding the set of tasks he is going to execute). Tasks are (i) \emph{location-dependent}, each associated with one or multiple target location(s) at which the task will be executed, and (ii) \emph{time-sensitive}, each associated with a valid time period within which the task must be executed.

Moreover, users are {geographically dispersed} (i.e., each associated with an initial location) and can travel to different locations for executing different tasks.
As different tasks may have different valid time periods, each user needs to decide not only the \emph{task selection} (i.e., the set of tasks he is going to execute) but also the \emph{execution order} of the selected tasks.
This is  also the key difference between the task scheduling in our work and that in \cite{ups-3}, which focused on the task selection only, without considering the execution order.\footnote{In \cite{ups-3}, each task is associated with a particular time, hence the execution order is inherently given as long as the tasks are selected.}
Note that our task scheduling problem (i.e.,   task selection and order optimization) is much more challenging than that  in \cite{ups-3} (i.e., task selection only), as even if the task selection is given, the execution order optimization is still an NP-hard problem.


Figure \ref{fig:mcs-model} illustrates an example of such a task scheduling decision in a UPS model with location-dependent time-sensitive tasks.
Each route denotes the task scheduling decision (i.e., task selection and execution order) of each user.
For example, user $1$ chooses to execute tasks $\{1,2,3,4\}$ in order, user $2$ chooses to execute tasks $\{5,1,6\}$ in order, and user 3 chooses to execute tasks $\{7,8,9\}$ in order.

\emph{Game Formulation}:
When a user executes a task successfully (i.e., at the   target location and within the   valid time period of the task), the user will obtain a certain \emph{reward} provided by the task owner.
When multiple users execute the same task, they will share the reward {equally} as in \cite{ups-3}.
This makes the task scheduling decisions of different users coupled with each other, leading to a \emph{strategic game} situation.

We formulate such a game, called \emph{Task Scheduling Game (TSG)}, and perform a comprehensive game-theoretic analysis.\footnote{Game theory  has been widely used in wireless networks (e.g., \cite{gao-1,gao-2,gao-3,gao-4,gao-5,gao-6}) for modeling and analyzing the competitive and cooperative interactions among different network entities.}
Specifically, we first prove that the TSG game is a  {potential game} \cite{potential}, which guarantees the existence of Nash equilibrium (NE).
Then we analyze the social efficiency loss at the NE (comparing with the socially optimal solution) induced by the selfish behaviors of users.
We further show how the efficiency loss changes with the user number and user type.

In summary, the main results and  key contributions of this work are summarized as follows.

\begin{itemize}
\item \emph{General Model:}
We consider a general UPS model with location-dependent time-sensitive tasks, which generalizes the existing task models in the literature.

\item \emph{Game-Theoretic Analysis:}
We perform a comprehensive game-theoretic analysis for the task scheduling   in the proposed UPS model, by using a potential game.

\item \emph{Performance Evaluation:}
We evaluate the efficiency loss and the fairness index at the NE under different situations.
Our simulations in practical scenarios with different types of users (walking, bike, and driving users) show that the efficiency loss can be up to $70\%$ due to the selfish behaviors of users.

\item \emph{Observations and Insights:}
Our analysis shows the NE performance may increase or decrease with the number of users, depending on the level of competition.
This implies that it is not always better to employ more users in the UPS system, which can provide a guidance for the system
designer to determine the optimal number of users to be
employed in a practical system.

\end{itemize}

The rest of the paper is organized as follows.
In Section \ref{section:model}, we present the system model.
In Section \ref{section:game} and \ref{section:analysis}, we formulate the task scheduling game and analyze the Nash equilibrium.
We present the simulation results in Section \ref{section:simulation}, and finally conclude in Section \ref{section:conclusion}.



%

\section{System Model}\label{section:model}

We consider a user-centric UPS system consisting of a sensing platform and a set $\N = \{1,\cdots,N\}$ of mobile smartphone users.
The platform announces a set $\S = \{1,\cdots,S\}$ of sensing tasks.
Each task can represent a specific sensing event at a particular time and  location, or a set of periodic sensing events within a certain time period, or a set of sensing events at multiple locations.
Each task $k \in \S$ is associated with a \emph{reward} $V_k$, denoting the money to be paid to the users who execute the task successfully.
Each user can choose the set of tasks he is going to execute.
When multiple users execute the same task, they will share the reward \emph{equally} as in \cite{ups-3}.
This makes the task scheduling decisions of different users coupled with each other, resulting in a  \emph{strategic game} situation.


\subsection{Task Model}
We consider a general task model, where tasks are (i)  \emph{location-dependent}: each task $k \in \S$ is associated with a target location $L_k $ at which the task will be executed;\footnote{Note that our analysis can be easily extended to the task model with multiple target locations, by simply dividing each task into multiple sub-tasks, each associated with one target location.}
and (ii) \emph{time-sensitive}: each task $k \in \S$ is associated with a valid time period $T_k \eq [T_k^{\dag},\ T_k^{\ddag}]$ within which the task must be executed.
Examples of such tasks includes the measurement of traffic speed at a particular road conjunction or the air quality of a particular location within a particular time interval.~~~~~~~~~~~~~~~~~

When enlarging the valid time period of each task to infinity, our model will degenerate to those in \cite{ups-0,ups-1,ups-2};
when shrinking the valid time period of each task to a single point, our model will degenerate to the model in \cite{ups-3}.
Thus,  our model generalizes the existing models in \cite{ups-0,ups-1,ups-2,ups-3}.
%



\subsection{User Model}

Each user $i \in \N $ can choose one or multiple tasks (to execute) from a set of tasks $\S_i \subseteq \S$ available to him.
The availability of a task to a user depends on factors such as the user's device capability, time availability, mobility, and experience.
When executing a task successfully, the user can get the task reward solely or share the task reward equally with other users who also execute the task.
The \emph{payoff} of each user is defined as the difference between the achieved \emph{reward} and the incurred \emph{cost}, mainly including the execution cost and  the travelling cost (to be described below).

When executing a task, a user needs to consume some time and device resource (e.g., energy, bandwidth, and CPU cycle), hence incur certain \emph{execution cost}.
Such an execution cost and time depends on both the task natures (e.g., one-shot or periodic sensing) and the user characteristics  (e.g., experienced or inexperienced, resource limited or adequate).
Let $T_{i, k}$ and $C_{i, k}$ denote the time and cost of user $i \in \N $ for executing task $ k \in \S_i $.
Each user $i$ has a total budget $C_i$ of resource that can be used for executing tasks.

Moreover, in order to execute a task, a user needs to move to the target location of the task (in a certain travelling speed), which may incur certain \emph{travelling cost}.
After executing a task, the user will stay at that location (to save travelling cost) until he starts to move to a new location to execute a new task.
By abuse of notation, we denote $L_i $ as the initial location of user $i\in \N$.
The travelling cost and speed mainly depend on the type of transportation that the user takes.
For example, a walking user has a low  speed and cost, while a driving user may have a high speed and cost.
Let  $\widetilde{C}_i$ and $R_i$    denote the travelling  cost (per unit of travelling distance) and speed (m/s) of user $i\in \N$, respectively.



\subsection{Problem Description}

As different tasks may have different valid time periods,
each user needs to consider not only the \emph{task selection} (i.e.,
the set of tasks to be executed) but also the \emph{execution order}  of
the selected tasks. As shown in Figure \ref{fig:mcs-model}, the task execution
order is important, as it affects not only the user's travelling
cost but also whether the selected tasks can be executed within
their valid time periods. For example, if user $1$ executes task
$4$ first (within the time period [13:00, 14:00], say 13:30), he
cannot execute tasks $\{1,2,3\}$ any more within their valid time
periods (all of which are earlier than 13:30). {This is also one of the key differences between our problem and
that in \cite{ups-3}, which only focused on the task selection, without
considering the task execution order.}

\section{Game Formulation}\label{section:game}

As mentioned before, when multiple users choose to execute the same task, they will share the task reward \emph{equally}.
This makes the decisions of different users coupled with each other, leading to a  \emph{strategic game} situation. In this section, we will provide the formal definition for such a game.

\subsection{Strategy and Feasibility}

As discussed in Section \ref{section:model}, the strategy of each user $i\in \N$ is to choose a set of available tasks (to execute) and the execution order of the selected tasks, aiming at maximizing his payoff.
Such a strategy of user $i$ can be formally characterized  by an \emph{ordered} task set, denoted by
\begin{equation}
\bs_i \eq \{ k_i^1, \cdots, k_i^{|\bs_i|} \} \subseteq \S_i
\end{equation}
where the $j$-th element $k_i^j $ denotes the $j$-th task selected and executed by user $i$.

A strategy $\bs_i = \{ k_i^1, \cdots, k_i^{|\bs_i|} \}  $ of user $i$ is feasible, only if the time-sensitivity constraints of all selected tasks in $\bs_i$ are satisfied,
or equivalently, there exists a reasonable execution time vector such that all selected tasks in $\bs_i$ can be executed within their valid time periods.
Let $[T_i^1,\cdots,T_i^{|\bs_i|}]$ denote a potential execution time vector of user $i$, where $T_i^j$ denotes the execution time for the $j$-th task $k_{i}^j$.
Then, $[T_i^1,\cdots,T_i^{|\bs_i|}]$ is feasible, only if (i) it satisfies the time-sensitivity constraints of all selected tasks, i.e.,
\begin{equation}\label{eq:mcs-fs1}
T_{k_i^j}^{\dag} \leq  T_i^j  \leq  T_{k_i^j}^{\ddag}, \quad j=1,\cdots, |\bs_i|,
\end{equation}
and meanwhile (ii) it is
reasonable in the temporal logic, i.e.,
\begin{equation}\label{eq:mcs-fs2}
\left\{
\begin{aligned}
T_i^{1} & \textstyle
 \geq \frac{D({i} ,  {k_i^{1}})}{R_i},
\\
T_i^{j } &\textstyle \geq T_i^{j-1} + T_{i, k_i^{j-1}} + \frac{D( {k_i^{j-1}}, {k_i^j}) }{R_i}, \  j = 2,\cdots, |\bs_i| ,
\end{aligned}
\right.
\end{equation}
where $D({i} ,  {k_i^{1}}) = |L_{i}- L_{k_i^{1}}| $ denotes the distance between user $i$'s initial location $L_i$ and the first task $k_i^{1}$, $D({k_i^{j-1}}, {k_i^{j}}) = |L_{k_i^{j-1}}- L_{k_i^{j}}| $ denotes the distance between tasks $k_i^{j-1}$ and $k_i^{j}$, and $T_{i, k_i^{j-1}}$ denotes the time for executing task $k_i^{j-1}$.\footnote{The first condition denotes that user $i$ needs to take at least the time $\frac{D({i} ,  {k_i^{1}})}{R_i}$ to reach the first task $k_i^{1}$,
and the following conditions mean that user $i$ needs to takes at least the time $T_{i, k_i^{j-1}} + \frac{D( {k_i^{j-1}}, {k_i^j}) }{R_i}$ to reach task $k_i^{j}$, where the first part of time is used for completing the previous task $k_i^{j-1}$ and the second part of time is used for travelling from task $k_i^{j-1}$ to task $k_i^{j}$.}

Moreover, a strategy   $\bs_i  $ of user $i$  is feasible, only if it satisfies the  resource budget constraint. That is,
\begin{equation}\label{eq:mcs-fs3}
\textstyle
\sum \limits_{k \in \bs_i} C_{i, k} \leq C_i.
\end{equation}

Based on the above, we express in the following lemma the feasibility conditions for the user strategy $\bs_i $.

\begin{lemma}[Feasibility]\label{lemma:mcs-feasibility}
A strategy $\bs_i $ of user $i$ is feasible, if and only if the conditions \eqref{eq:mcs-fs1}-\eqref{eq:mcs-fs3} are satisfied.
\end{lemma}

Intuitively, if one of the conditions in \eqref{eq:mcs-fs1}-\eqref{eq:mcs-fs3} is not satisfied, then the strategy $\bs_i $ is not feasible as explained early. Thus,  if $\bs_i $ is feasible, the conditions  \eqref{eq:mcs-fs1}-\eqref{eq:mcs-fs3} must be satisfied.
One the other hand, if \eqref{eq:mcs-fs1}-\eqref{eq:mcs-fs3} are satisfied, the strategy $\bs_i $ can be implemented  successfully, hence is feasible.


\subsection{Payoff Definition}

Given a feasible strategy profile $\bs \eq (\bs_1,\cdots,\bs_N)$, i.e., the feasible strategies of all users,
we can compute the number of users executing task $k \in \S$, denoted by $M_k (\bs)$, that is,
\begin{equation}\label{eq:xxx:mk}
\textstyle
M_k (\bs) = \sum\limits_{i \in \N} \mathbf{1}_{(k \in \bs_i)}, \quad \forall k\in \S,
\end{equation}
where the indicator    $\mathbf{1}_{(k \in \bs_i)} = 1$ if $ k \in \bs_i$, and $  0$ otherwise.
Then, the total reward of each user $i \in \N $ can be computed~by
\begin{equation}\label{eq:xxx:reward}
\textstyle
r_i (\bs) \eq r_i (\bs_i, \bs_{-i}) = \sum\limits_{k \in \bs_i} \frac{V_k}{M_k (\bs)} ,
\end{equation}
which depends on both his own strategy $\bs_i$ and the strategies of other users, i.e., $\bs_{-i} \eq  (\bs_1,\cdots,\bs_{i-1},\bs_{i+1},\cdots,\bs_N) $.
The total execution cost of user $i \in \N $ can be computed by
\begin{equation}\label{eq:xxx:excost}
\textstyle
  c_i^{\textsc{ex}} (\bs_i ) = \sum \limits_{k \in \bs_i} C_{i,k},
\end{equation}
which depends  only on his own strategy  $\bs_i$.
The total travelling cost of user $i \in \N$ can be computed by
\begin{equation}\label{eq:xxx:excost}
\textstyle
  c_i^{\textsc{tr}} (\bs_i ) =     \sum\limits_{j = 1}^{ |\bs_i|  } D({k_i^{j-1}}, {k_i^{j}}) \cdot \widetilde{C}_{i}  ,
\end{equation}
which depends only on his own strategy  $\bs_i$.
Here we use the index $k_i^0$ to denote user $i$'s initial location.
Based on the above, the payoff of  each user $ i \in \N $ can be written as follows:
\begin{equation}\label{eq:xxx:payoff}
\begin{aligned}
\u_i (\bs) \eq \u_i (\bs_i, \bs_{-i})
  & = r_i (\bs  ) - c_i^{\textsc{ex}} (\bs_i ) - c_i^{\textsc{tr}} (\bs_i )
\\
& \textstyle = \sum \limits_{k \in \bs_i} \frac{V_k}{M_k (\bs)} - c_i^{\textsc{ex}} (\bs_i ) - c_i^{\textsc{tr}} (\bs_i ).
\end{aligned}
\end{equation}

\subsection{Task Scheduling Game -- TSG}
Now we define the Task Scheduling Game (TSG) and the
associated Nash equilibrium (NE) formally.

\begin{definition}[\textbf{Task Scheduling Game -- TSG}]\label{def:task-game}
The Task Selection Game  $\T \eq (\N, \{\S_i\}_{i\in\N}, \{\u_i\}_{i\in\N})$ is defined by:~~~~~~~~~
\begin{itemize}
  \item  \textbf{Player}: the set of participating users $\N = \{1, \cdots , N\}$;
  \item  \textbf{Strategy}: an ordered set of available tasks $\bs_i \subseteq \S_i$ for each participating  user $i\in\N$;
  \item  \textbf{Payoff}: a payoff function   $ \u_i (\bs_i, \bs_{-i})$ defined in \eqref{eq:xxx:payoff} for each participating  user $i\in\N$.
\end{itemize}
\end{definition}

A feasible strategy profile $\bs^* \eq (\bs_1^*,\cdots,\bs_N^*)$ is an NE of the Task Scheduling Game $\T$, if
 \begin{equation}\label{eq:xxx:nene}
\begin{aligned}
 \bs_i^* = \arg \max_{\bs_i \subseteq \S_i} \  & \u_i (\bs_i, \bs_{-i}^*) \\
  s.t.\  & \bs_i \mbox{ satisfies \eqref{eq:mcs-fs1}-\eqref{eq:mcs-fs3},}
\end{aligned}
\end{equation} 
for every user $i\in\N$.

\section{Game Equilibrium Analysis}\label{section:analysis}

We now analyze the NE of the Task Scheduling Game.

\subsection{Potential Game}

We first show that the Task Scheduling Game $\T$ is~a~\emph{potential game} \cite{potential}.
A game is called an (exact) potential game, if there exists an (exact) \emph{potential function}, such that for any user, when changing his strategy, the change of his payoff is equivalent to that of the potential function.
Formally,
\begin{definition}[Potential Game \cite{potential}]\label{def:potential-game}
A game $\G = (\N,$ $\{\S_i\}_{i\in\N},\ \{\u_i\}_{i\in\N})$ is called a potential game, if it admits a potential function $\Phi(\bs)$ such that for every player $i\in \N$ and any two strategies $\bs_i,\bs_i' \subseteq \S_i $ of player $i$,
\begin{equation}\label{eq:pf-proof}
\u_i (\bs_i, \bs_{-i} ) - \u_i (\bs_i', \bs_{-i} )
=
\Phi (\bs_i, \bs_{-i} ) - \Phi (\bs_i', \bs_{-i} ),
\end{equation}
under any strategy profile $\bs_{-i}$ of players other than $i$.
\end{definition}

\begin{lemma}\label{lemma:mcs:potential}
The Task Scheduling Game $\T$ is a potential game, with the following potential function  $\Phi(\bs)$:
\begin{equation}\label{eq:xxx:pf}
\begin{aligned}
\textstyle
 \Phi(\bs) =  \sum\limits_{k \in \S} \sum \limits_{m=1}^{M_k(\bs)} \frac{V_k   }{m}
  - \sum \limits_{i \in \N } c_i^{\textsc{ex}} (\bs_i ) - \sum \limits_{i \in \N } c_i^{\textsc{tr}} (\bs_i ),
\end{aligned}
\end{equation}
where $M_k(\bs)$ is defined in \eqref{eq:xxx:mk}, i.e., the total number of users executing task $k$ under the strategy profile $\bs$.
\end{lemma}

The above lemma can be proved by showing that the condition  \eqref{eq:pf-proof} holds under the following two situations.
First, a user changes the execution order, but not the task selection.
Second, a user changes the task selection by adding an additional task or removing an existing task.
Due to space limit, we put the detailed proof in our online technical report \cite{report}.

\begin{figure*}
\hspace{-5mm}
	\centering
\includegraphics[width=2.3in]{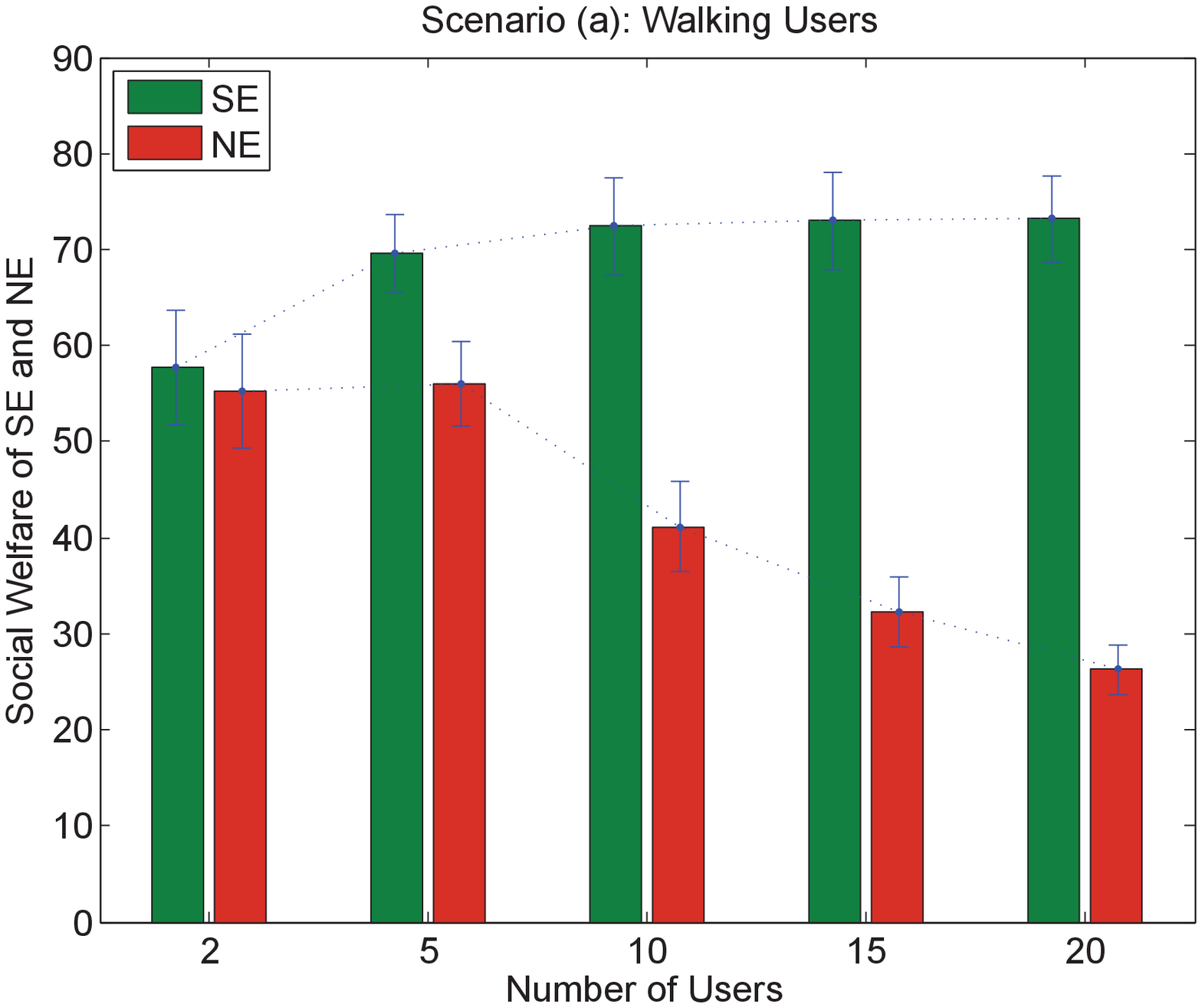}
~~
~~
\includegraphics[width=2.3in]{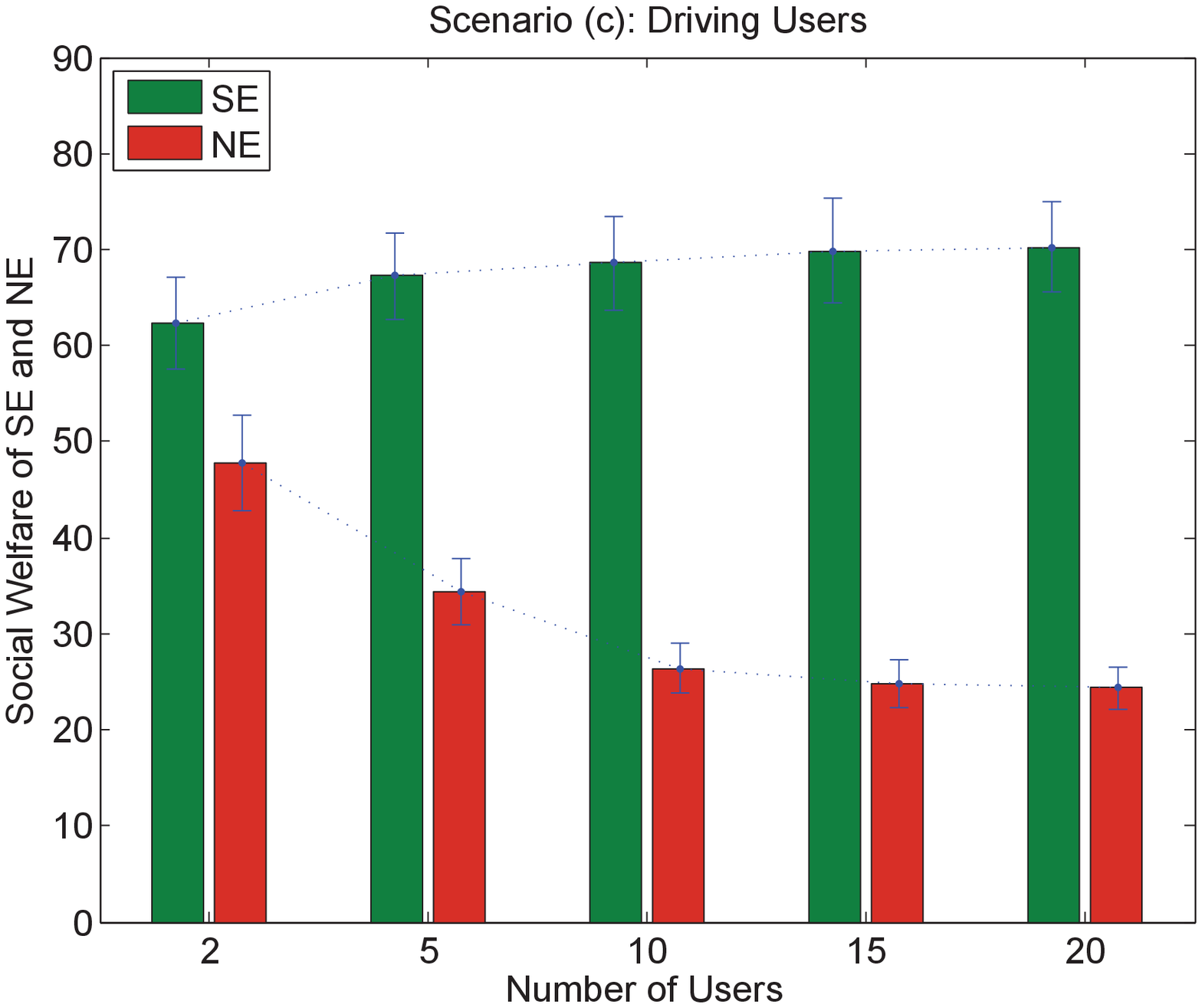}
\vspace{-3mm}
	\caption{Social Welfare under SE and NE: (a) Walking Users; (b) Bike Users; (c) Driving Users.}
\label{fig:mcs-simu-sw}
\vspace{-3mm}
\end{figure*}

\subsection{Nash Equilibrium -- NE}

We now analyze the NE of the proposed game. As shown in \cite{potential},
an appealing property of a potential game is that it
always admits an NE. In addition, any strategy profile $\bs^*$ that maximizes the potential function $\Phi(\bs)$ is an NE. Formally,
 \begin{lemma}\label{lemma:mcs:NE}
The Task Scheduling Game $\T$ has at least one NE $\bs^* \eq (\bs_1^*,\cdots,\bs_N^*)$, which is  given by
\begin{equation}\label{eq:xxx:NE}
\begin{aligned}
 \bs^* \eq  \arg \max_{\bs} &\  \  \Phi(\bs)
 \\
   s.t.\  & \bs_i \mbox{ satisfies \eqref{eq:mcs-fs1}-\eqref{eq:mcs-fs3}}, \ \forall i\in \N,
\end{aligned}
\end{equation}
where $\Phi(\bs)$ is the potential function defined in \eqref{eq:xxx:pf}.
\end{lemma}

This lemma can be easily proved by observing that
$$
 \u_i (\bs_i^*, \bs_{-i}^* ) - \u_i (\bs_i', \bs_{-i}^* )
=
\Phi (\bs_i^*, \bs_{-i}^* ) - \Phi (\bs_i', \bs_{-i}^* ) \geq 0,
$$
for any user  $i\in \N$ and   $\bs_i ' \subseteq \S_i$. The last inequality follows because $(\bs_i^*, \bs_{-i}^*)$ is the maximizer of
$\Phi (\bs)$ by \eqref{eq:xxx:NE}.

\subsection{Efficiency of NE}

Now we show that the NE of the Task Scheduling Game
$\T$, especially the one given by \eqref{eq:xxx:NE}, is often \emph{not} efficient.

Specifically, a strategy profile $\bs^\circ$ is socially efficient (SE), if it maximizes the following social welfare:
\begin{equation}\label{eq:xxx:sw}
\textstyle
\W(\bs) = \sum\limits_{k \in \S} V_k  \cdot \big( \mathbf{1}_{ M_k(\bs) \geq 1 } \big)
 - \sum \limits_{i \in \N } \big( c_i^{\textsc{ex}} (\bs_i ) + c_i^{\textsc{tr}} (\bs_i ) \big) ,
\end{equation}
where $\mathbf{1}_{ M_k(\bs) \geq 1 } = 1$ if $M_k(\bs) \geq 1$, and $0$ otherwise. The first term denotes the total reward collected by all users, where the reward $V_k$ of a task $k$ is collected if at least one user   executes the task successfully (i.e., $M_k(\bs) \geq 1$).
The second term denotes the total cost incurred on all users.
Formally, the socially efficient solution $\bs^\circ$ is given by:
\begin{equation}\label{eq:xxx:soooo}
\begin{aligned}
 \bs^\circ \eq  \arg \max_{\bs} &\  \  \W(\bs)
 \\
   s.t.\  & \bs_i \mbox{ satisfies \eqref{eq:mcs-fs1}-\eqref{eq:mcs-fs3}}, \ \forall i\in \N.
\end{aligned}
\end{equation}

By comparing  $\W(\bs) $ in \eqref{eq:xxx:sw}
and
 $ \Phi(\bs)$ in \eqref{eq:xxx:pf}, we can see that both functions have the similar structure, except the
coefficients for $ V_k, k\in \S $ in the first term, i.e.,
$
\sum_{m=1}^{M_k(\bs)} \frac{1  }{m}
$
and
$   \mathbf{1}_{( M_k(\bs) \geq 1 )}
$.
We can further see that
$$
\textstyle
\sum\limits_{m=1}^{M_k(\bs)} \frac{1  }{m}
\geq    \mathbf{1}_{( M_k(\bs) \geq 1 )},
$$
where the equality holds only when $M_k(\bs) = 1$ (both sides are $1$) or $0$ (both sides are $0$).
Namely, the coefficient for each $ V_k$  in $ \Phi(\bs) $ is no smaller than that in $\W(\bs) $.
This implies that for any task, users are more likely to execute the task at the NE than SE.
This leads to the following observation.
\begin{observation}
The task selections at the NE, especially that resulting from  \eqref{eq:xxx:NE}, are more  {aggressive}, comparing with those at the SE resulting from maximizing $ \W(\bs) $.
\end{observation}


\ifodd 0
To illustrate this, we provide a simple example with one task and two users. Suppose that the reward of the task is $10$, the travelling costs of both users are zero, and the sensing costs of users are $4.8$ and $4.9$, respectively.
The SE outcome is: only user 1  executes the task, leading to a social welfare of $10-4.8 = 5.2$.
Under the NE, however, both users will choose to execute the task (as both can achieve a positive payoff even if they share the reward), leading to a social welfare of $10-4.8-4.9 = 0.3$.

\begin{figure}
	\centering
\includegraphics[width=2.8in]{Figures/simu-1new}
\vspace{-2mm}
	\caption{Social Welfare under SE and NE.}
\label{fig:simu-1}
\vspace{-4mm}
\end{figure}

In order to obtain more useful engineering insights, we also perform numerical studies to illustrate the efficiency loss at the NE (comparing with the SE) under different situations.
Figure \ref{fig:simu-1} illustrates the normalized social welfare under the NE (the blue bars) and the SE (the hollow bars with black borders).
Each bar group denotes the results under a particular task reward (e.g., $V_k = 0.2$ for the first group and $V_k = 1$ for the last group), and different bars in the same group correspond to different numbers of users (e.g., $N=2$ for the first bar and $N=14$   for the last bar in each group).
From Figure \ref{fig:simu-1}, we have the following observations.
\begin{observation}\label{ob:2}
The social welfare under SE increases with the number of users (for any possible task reward).
      \emph{The reason is that with more users, it is more likely to choose  users with smaller costs to execute the tasks.}
 \end{observation}

\begin{observation}\label{ob:3}
When the task reward is small (e.g., $V_k = 0.2$), the social welfare under NE   increases with the number of users.
\emph{The reason is similar as that for SE in Observation \ref{ob:2}: with more users, it is more likely to choose users with  smaller costs to execute  the tasks.}
\end{observation}

\begin{observation}\label{ob:5}
When the task reward is large (e.g., $V_k  = 1$), the social welfare under NE decreases with the number of users.
\emph{The reason is that with more users it is more likely to choose  multiple users to execute the same task, which will reduce the social welfare.}
\end{observation}

\begin{observation}\label{ob:4}
When the task reward is medium (e.g., $V_k \in [0.4,0.8]$), however, the social welfare under NE first increases and then decreases with the number of users.
\emph{The reason is similar as those in Observations \ref{ob:3} and \ref{ob:5}.}
\end{observation}


\fi

\section{Simulations}\label{section:simulation}


\subsection{Simulation Setting}
We choose a 5km$\times$5km region in London as the simulation area, where tasks and users are randomly  distributed in the area.
We simulate a time period of $2$ hours, e.g., [10:00, 12:00], within which each task is initiated randomly and uniformly.
The task valid (survival) time of each task is selected randomly according to the (truncated) normal distribution with the expected value of $30$ minutes.
Each task needs to be started (but not necessarily to be completed) within its valid time period.
The reward of each task is selected randomly according to  the (truncated) normal distribution with the expected value of $\$10$ (in dollar).
In the simulations, we fix the number of tasks to $8$, while change the number of users from $2$ to $20$ (to capture different levels of competition among users).

We consider three types of different users according to their travelling modes: (i) \emph{Walking users}, who travel by walking, with a relatively low travelling speed $5$km/h and cost $\$0.2$/km; (ii) \emph{Bike users}, who travel by bike, with a medium travelling speed $15$km/h and cost $\$0.5$/km, and (iii) \emph{Driving users}, who travel by driving, with a relatively high travelling speed  $45$km/h  and cost $\$1$/km.
Each user takes on average $10$ minutes to execute a task, incurring an execution cost of $\$1$ (in dollar) on average.
Both the execution time and cost follow the (truncated) normal distribution as other parameters.

\subsection{Social Welfare Gap}
We first show the social welfare gap between the NE and SE, which captures the efficiency loss of the NE.

Figure \ref{fig:mcs-simu-sw} presents the expected social welfare under SE and NE with three different types of users.
In the first subfigure (a),  users are walking users with low travelling speed and cost.
In the second subfigure (b), users are bike users with medium travelling speed and cost.
In the third subfigure (c), users are driving users with high travelling speed and cost.
From Figure \ref{fig:mcs-simu-sw}, we have the following observations:

\emph{1)}
\emph{The social welfare under SE always increases with the number of users in all three scenarios (with walking, bike, and driving users).}
This is because with more users, it is more likely that tasks can be executed by lower cost users, hence resulting in a higher social welfare.

\emph{2)}
\emph{The social welfare under NE decreases with the number of users in most cases.}
This is because with more users, the competition among users becomes more intensive, and the probability that multiple users choosing the same task increases, which will lead to a higher total execution/travelling cost and hence a lower social welfare.

\emph{3)}
\emph{The social welfare under NE may also increase with the number of users in some cases.}
For example, in scenario (a) with walking users, when the number of walking user is small (e.g., less than 5), the social welfare under NE increases with the number of users slightly.
This is because the walking users often have a limited serving region due to the small travelling speed, hence there is almost no competition among users when the number of users is small.
In this case, increasing the number of users will not introduce much competition among users, but increase the probability that the tasks being executed by lower cost users, hence resulting in a higher social welfare.
\ifodd 0
This coincides with the numerical results in Figure \ref{fig:simu-1}, where the social welfare under NE  increases with the number of users when the user number is small.
\fi


\subsection{Social Welfare Ratio}

\begin{figure}
\vspace{-3mm}	\centering
\includegraphics[width=2.8in]{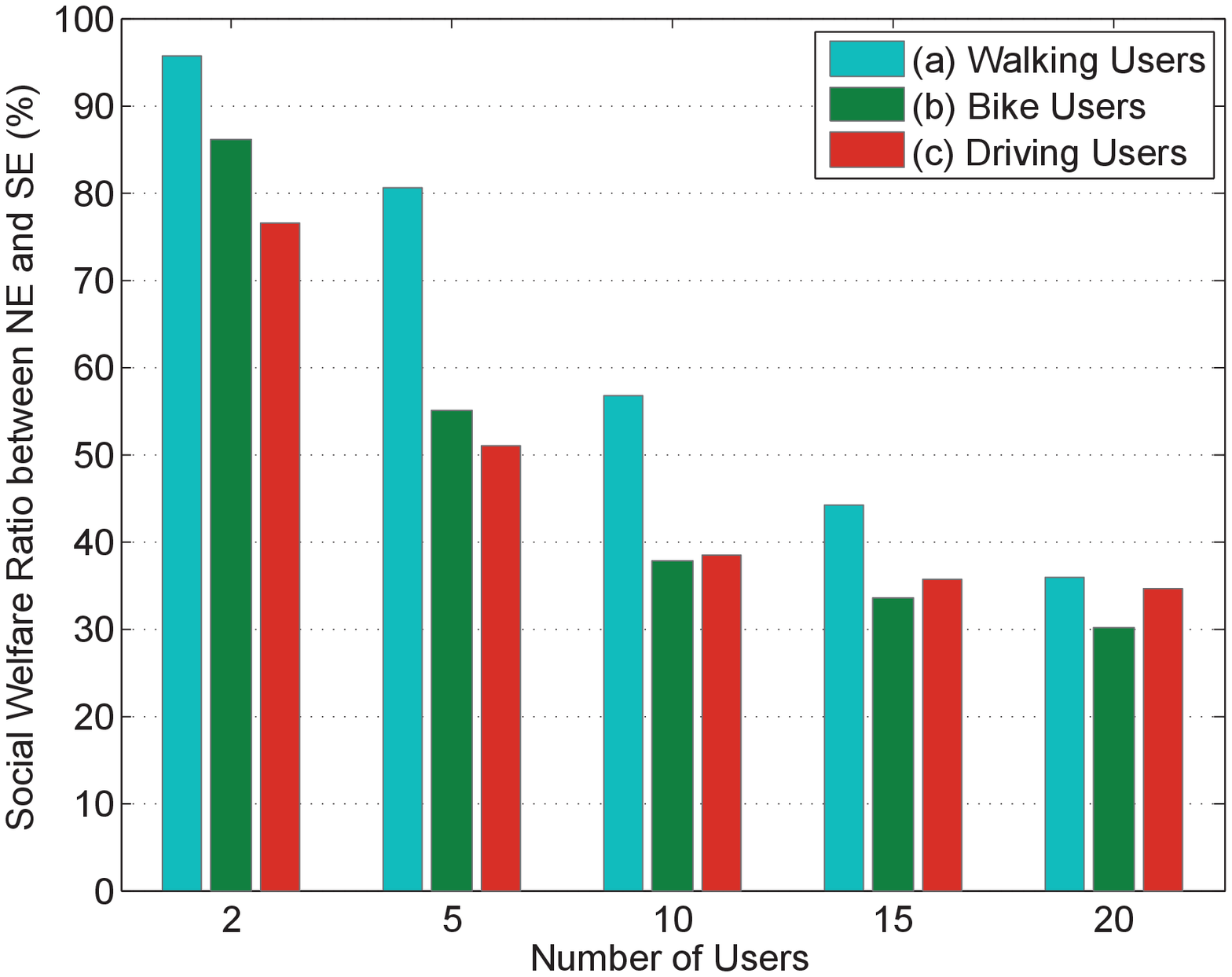}
\vspace{-3mm}
	\caption{Ratio of Social Welfare between SE and NE.}
\label{fig:mcs-simu-ratio}
\vspace{-3mm}
\end{figure}

Figure \ref{fig:mcs-simu-ratio} further  presents the social welfare \emph{ratio} between NE and SE with three different types of users.
We can see that in all three scenarios, the social welfare ratio decreases with the number of users.
This is because the social welfare increases with the number of users under SE, while (mostly) decreases with the number of users under NE, as illustrated in the previous Figure \ref{fig:mcs-simu-sw}.
We can further see that the social welfare ratio with walking users is higher than those with bike users and driving users.
This is because the competition among walking users is less intensive than that among bike/driving users, due to the limited traveling speed and serving region of walking users.
Hence, the degradation of social welfare under NE (comparing with SE) is smaller with walking users.
More specifically, we can find from Figure \ref{fig:mcs-simu-ratio}  that when the number of users changes from $2$ to $20$,  the social welfare ratio decreases from $95\%$, $85\%$, and $75\%$ to approximately $30\%$ for walking, bike, and driving users, respectively.

\ifodd 0

\subsection{Fairness}

We now show the fairness at the NE, which captures how the generated social welfare is distributed among users.
We evaluate the fairness by the widely-used Jain's fairness index \cite{Jain}, which is defined as follows:
$$
J(\bs) = \frac{\left( \sum_{i\in\N} \u_i(\bs) \right)^2}{N \cdot \sum_{i\in\N} \u_i(\bs)^2  } .
$$
By the above definition, it is easy to see that the maximum Jain's fairness is $1$, which can be achieved when all users share the social welfare equally, and the minimum Jain's fairness is $\frac{1}{N}$, which can be achieved when one user gets all of the social welfare while all other users get a zero payoff.

Figure \ref{fig:mcs-simu-jane} presents the Jain's fairness index under NE with three different types of users.
We can see that the Jain's fairness index decreases with the number of users, and can be down to $0.8$, $0.6$, and $0.48$ for walking, bike, and driving users, respectively, when the total number of users is 20.
This implies that a larger number of users will lead to a poor fairness under NE.
This is because a larger number of users will lead to a more intensive competition among users, hence it is more likely that some highly competitive users occupy most of the social welfare, resulting in a lower fairness index.

Figure \ref{fig:mcs-simu-jane} also shows that under NE, the Jain's fairness index with walking users is often higher than those with bike users and driving users.
This is because the competition among walking users is less intensive than that among bike or driving users, due to the limited traveling speed and serving region of walking users.
Thus, with bike or driving users, it is more likely that some highly competitive users occupy most of the social welfare, resulting in a lower fairness index.


\fi

\section{Conclusion}\label{section:conclusion}

In this work, we study the task scheduling in the user-centric
PS system by using a game-theoretic analysis. We formulate
the strategic interaction of users as a task scheduling game, and
analyze the NE by using a potential game. We further analyze
the efficiency loss  at the NE under
different situations. There are several interesting directions for
the future research. First, our analysis and simulations show
that the social efficiency loss at the NE can be up to 70\%.
Thus, it is important to design some mechanisms to reduce
the efficiency loss. Second, the current model did not consider
the different efforts of users in executing a task. It is important
to incorporate the effort into the user decision.



\end{document}